# A BIOMECHANICAL MODEL OF THE FACE INCLUDING MUSCLES FOR THE PREDICTION OF DEFORMATIONS DURING SPEECH PRODUCTION

Julie Groleau[1], Matthieu Chabanas[1], Christophe Marécaux[3], Natacha Payrard[2],
Brice Segaud[2], Michel Rochette[2], Pascal Perrier[1] and Yohan Payan[3]
[1]ICP-GIPSA Lab, UMR CNRS 5216, INP Grenoble, France
[2]ANSYS France, Villeurbanne, France
[3]TIMC-IMAG, UMR CNRS 5525, Université Joseph Fourier, Grenoble, France

*Abstract: A 3D biomechanical finite element model of the face is presented. Muscles are represented by piece-wise uniaxial tension cable elements linking the insertion points. Such insertion points are specific entities differing from nodes of the finite element mesh, which makes possible to change either the mesh or the muscle implementation totally independently of each other. Lip/teeth and upper lip/lower lip contacts are also modeled. Simulations of smiling and of an Orbicularis Oris activation are presented and interpreted. The importance of a proper account of contacts and of an accurate anatomical description is shown.*
*Keywords : Face models, Muscle modeling, Lip/teeth interaction.*

## I. INTRODUCTION

Many biomechanical models of the human face have been proposed in the literature. They were generally developed either in the context of computer graphics animation [1,2], or of computer-aided maxillofacial surgery [3,4] or of speech production studies[5]. Most of them propose to model the face with a volumetric mesh defined by an external (the "visible" part of the face) and an internal surface (the part in contact with the skull), with some nodes or layers in between. Mechanics of the tissues (epidermis, dermis, hypodermis, fat and muscles) is then modelled through the relation between displacements and forces of mass points, or through strain/stress relations in the case of finite element models. These studies have raised a number of important issues: (1) how to model muscles fibres and their action on the 3D mesh; (2) how to account for the subject/patient specific muscular morphology (in terms of fibres insertions and interweaving); (3) how to control the large number of muscles in order to produce a given speech articulation or facial mimics. Both latter points have already been addressed and discussed by our group, respectively for computer-aided craniofacial surgery [4] and through a motor control model of tongue muscles activation for speech production [6].

This paper deals with the first issue and proposes a method to define from Computer Tomography (CT) images a subset of muscles fibres within a 3D mesh of the face. Contacts between lips and teeth are also handled. First results of facial mimics' simulations are presented.

## II. METHODS

The starting point of this modeling work is the 3D Finite Element model of the face soft tissues, built out of CT scan of a single patient, which was originally proposed in [4]. It relies on a volumetric mesh consisting of hexahedrons and wedges elements (Fig. 1, left). The displacements of several nodes located on the internal surface of the facial mesh are constrained in order to represent attachments of the facial tissues on the skull.

While biological soft tissues are known to behave non-linearly [7], they were first represented by a homogeneous, isotropic, linear material. This hypothesis was retained in a first stage in order to focus on muscles modelling and contact management, before ongoing with more realistic modelling. Simulations were computed using the ANSYS[TM] v11 finite element software.

The first part of our study has consisted in building the muscles involved in facial mimics' generation. In order to ensure anatomical and physical reliability, muscles courses and insertions were directly defined from medical images and anatomical charts (Fig.1, middle), with the help of a maxillofacial surgeon. The locations of points describing the muscle fibres were measured in the different scan slices. These points were then integrated into the mesh to model muscle insertions. They were linked with piece-wise uniaxial tension cable elements to model muscle fibres (Fig. 1, right). The Orbicularis Oris muscle was designed slightly differently: it is represented by two ellipsoid cable elements centred on the mouth opening, and representing the marginal and the peripheral parts of the muscle.

The cable elements based approach allows integrating muscles into the model independently of the mesh itself. Therefore, the mesh can be easily refined or modified, without requiring any change in the muscles structure definition. The fibers cable elements are controlled in tension by their cross section area, their initial strain and an activation parameter. They generate forces that are applied to the soft tissues mesh thanks to the notion of dependencies. In other words, muscular fibres extremities are linked with the facets of the surrounding mesh elements. When a muscle is activated, the corresponding cable elements exert forces on the mesh elements and induce, then, soft tissue deformations.

The second part of our study concerns lips-teeth and upper-lip/lower-lip contacts, which are of primary importance in lips movements and deformations. Teeth are materialized in the model by surfaces extracted from the CT data and interpolated with Spline functions. ANSYS contact elements, which provide collision detection and sliding reaction, are used to mesh lips and teeth surfaces.

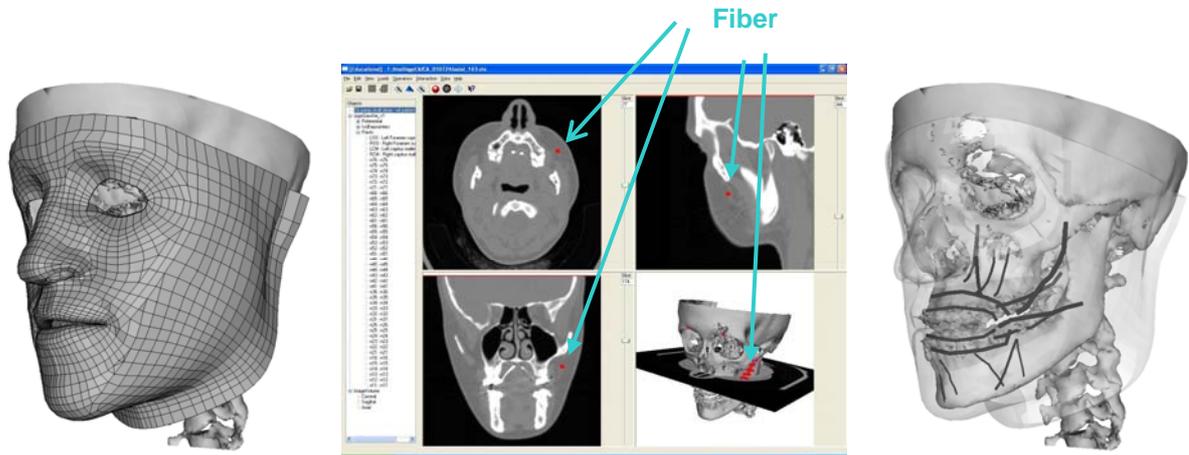

**Figure 1** Left: the finite element mesh of the face soft tissues. Middle: interactive segmentation of muscles fibers on CT data. Right: location of eleven muscles involved in facial mimics on the left side of the skull.

III. RESULTS

**a. Simulating smiling lips.**

Figure 2 presents the mesh deformations (in colours/grey scale) and the final face shape when the Zygomaticus Major, the Risorius and the Levator Labii Superioris are activated simultaneously. The bottom panels show the results when lip/lip and lip/teeth contacts are taken into account, while the top panels show the results obtained without contact. In presence of contacts, the mimic seems more realistic. This can be particularly well assessed on the side views (right panel), where, in the absence of contacts, an interpenetration of the lips can be noticed. On the contrary,, lips are slightly opened in the other condition, which is in agreement with data on smiling.

**b. Orbicularis activation and lips protrusion.**

An important issue for speech production concerns the control of protruded lips, such as in the production of /u/ or /y/ in French. It has been suggested in the literature [8] that the interaction between an activation of the Orbicularis Oris and lip/lip and lip/teeth contacts could be responsible for this particular lip gesture. This hypothesis was tested with our face model.

Two types of simulations were then run with the activation of the Orbicularis Oris, without (Figure 3 top panels) and with (Figure 3 low panels) handling contacts. In the absence of contacts, a rounding of the lips is observed (top panel, left). Rounding is classically associated with protrusion. However, the side view (top panel, right) shows a strong retraction of the lips, which is at the opposite of a protrusion. Including the contact limits the retraction, but it does not generate any protrusion or rounding.

The absence of protrusion can be explained by the fact that our model does not separately control the marginal and the peripheral parts of the Orbicularis Oris. Honda et al. [8] found namely an EMG activation during protrusion only in the peripheral part of the muscle. Gomi et al. [5] confirmed this finding with their biomechanical lip model. However, the absence of rounding is in agreement with Gomi et al.'s [5] statement (p.130) who suggested that not only the Orbicularis Oris, but also "*additional muscles combinations (jaw opening, peripheralis or other perioral muscles) would be required to form rounded lips.*" "

## IV. DISCUSSION AND CONCLUSION

The face model presented in this paper integrates an original representation of muscle fibres and muscle force generation in a 3D mesh based on piece-wise uniaxial tension cable elements. It also models contacts between upper lip and lower lip and between lips and teeth in a realistic way.

Simulations of smiling lips show that the proposed muscle representation is adapted to the generation of lip deformations that are realistic both in amplitudes and in directions. This is an important result since this representation can be implemented independently of the mesh. It will then facilitate the generation of speaker specific mesh using mesh matching algorithm [9]. It will also increase the efficiency of such a modelling approach to study the impact of face surgery on smiling and on mimics in general. Indeed, the geometrical structure of the mesh can easily be modified to account for different kinds of surgeries, without inducing a careful, difficult and long redefinition of all muscles fibres in the mesh.

Our results show also the importance of contacts modelling, both for lips/teeth interactions, but also for upper lip/lower lip interaction. It is important, not only because it prevents for unrealistic interpenetrations, but also because it allows sliding movements that constrain and guide the movement. This is certainly a phenomenon underlying complex lip shaping such as protrusion and rounding.

On the other hand, the simulations of the consequences of the activation of the Orbicularis Oris, for which no distinction is made in our model between the marginal and the peripheral parts, do not generate rounding. These results are in contradiction with the simulations carried out by Gomi et al's [5] who did make this distinction. This shows that collecting accurate neurophysiological and anatomical data is a major challenge to test hypotheses about the control of speech gestures, once realistic biomechanical models are available.
.

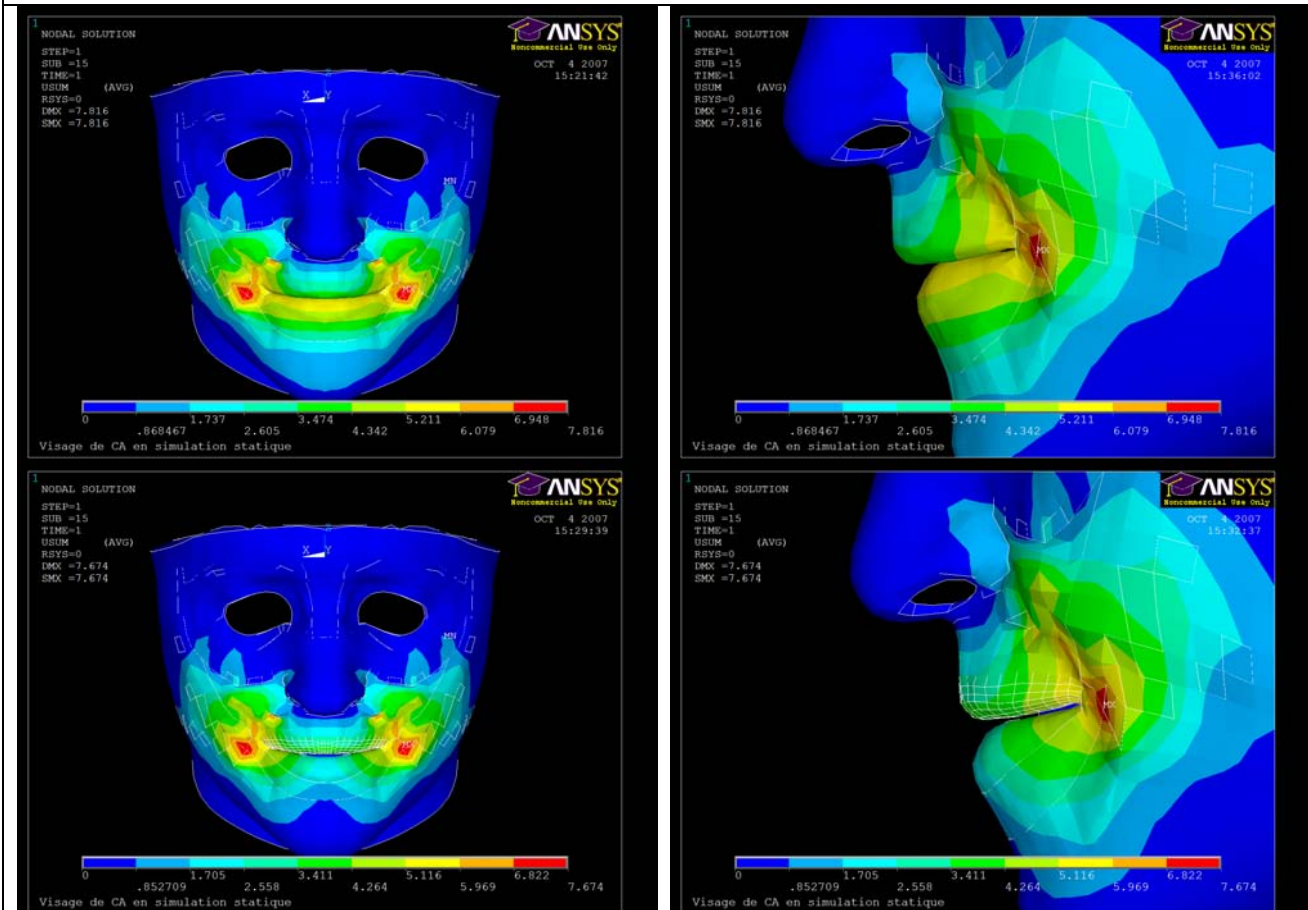

**Figure 2** Simulation of smiling lips without (top panels) and with (bottom panels) handling of lips-teeth and upper-lower lips contacts. Displacements are in mm.

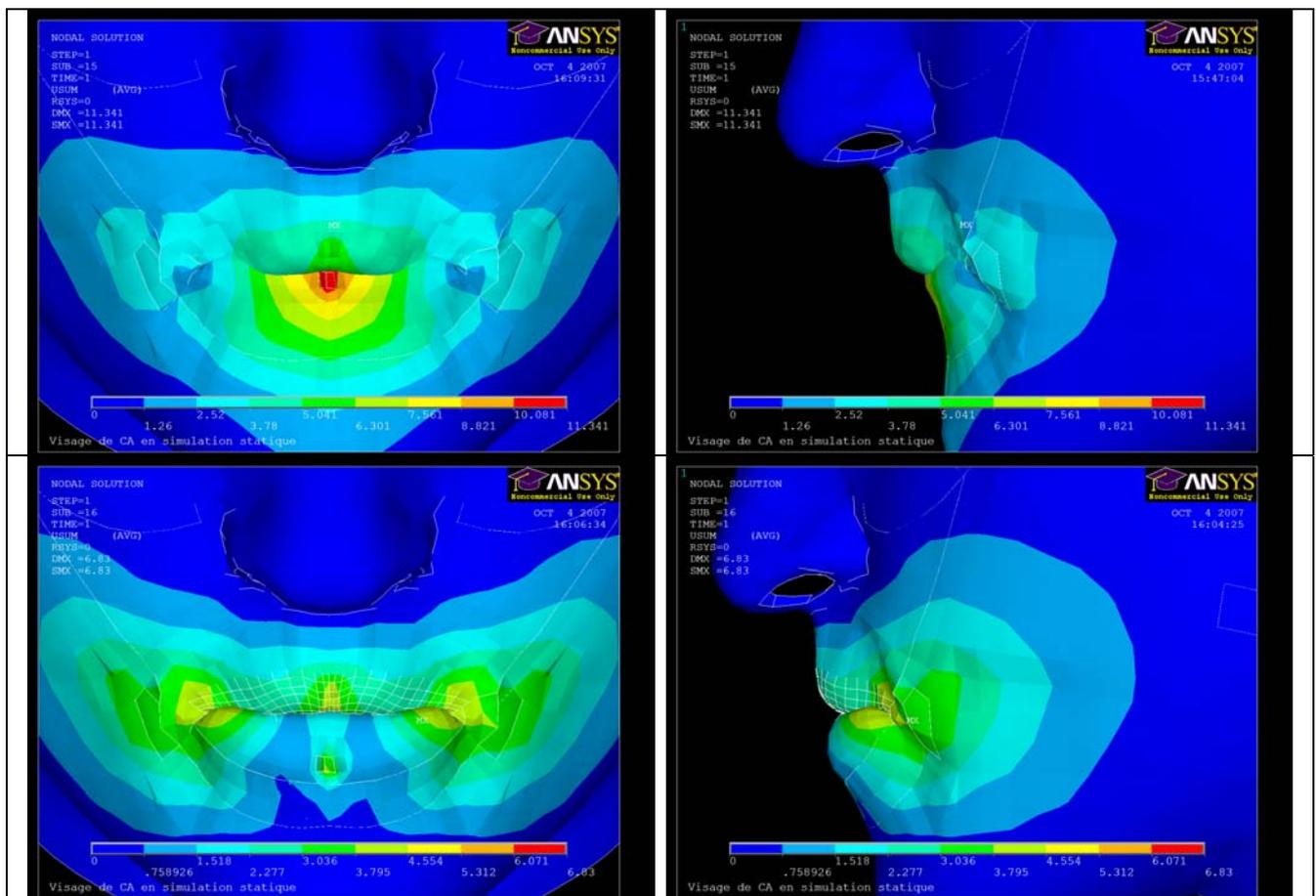

**Figure 3** Simulation of the Orbicularis Oris contraction, without (top panels) and with (bottom panels) handling of lips-teeth and upper-lip/lower-lip contacts.

REFERENCES

[1] Lucero, J.C. & Munhall, G.K. (1999). A model of facial biomechanics for speech production. *J. Acoustic Soc. Am.*, 106(5):2834-2842.

[2] Sifakis, E., Selle, A., Robinson-Mosher, A., Fedkiw, R. (2006). Simulating Speech with a Physics-Based Facial Muscle Model. ACM SIGGRAPH/*Eurographics Symposium on Computer Animation :*261-270

[3] Gladilin, E., Zachow, S., Deuflhard, P., Hege, H.C. (2001). Towards a Realistic Simulation of Individual Facial Mimics. *VMV:*129-134.

[4] Chabanas, M., Luboz, V. & Payan, Y. (2003). Patient specific Finite Element model of the face soft tissue for computer-assisted maxillofacial surgery, *Medical Image Analysis*, Vol. 7, Issue 2, pp. 131-151.

[5] Gomi, H, Nozoe, J, Dang, J, Honda, K. (2006). A physiologically based model of perioral dynamics for various lip deformations in speech articulation. In Harrington J. & Tabain M. (Eds.), *Speech Production-Models, Phonetic Processes, and Techniques.* Psychology Press, New-York, USA

[6] Buchaillard, S, Perrier, P & Payan, Y. (2006) A 3D biomechanical vocal tract model to study speech production control: How to take into account the gravity? *Proceedings of the 7th International Seminar on Speech Production* (pp.403-410), Ubatuba, Brazil.

[7] Gérard, J.-M., Ohayon, J., Luboz, V., Perrier, P. & Payan, Y. (2005). Non linear elastic properties of the lingual and facial tissues assessed by indentation technique. Application to the biomechanics of speech production. *Medical Engineering & Physics,27*, 884–892.

[8] Honda K.,Kurita T., Kakita Y., and Maeda S., Physiology of the lips and modeling of lip. gestures, *J Phonetics*, 23, 243-54

[9] Couteau, B., Payan, Y. & Lavallée, S. (2000). The Mesh-Matching algorithm: an automatic 3D mesh generator for finite element structures. *Journal of Biomechanics, 33(8)*, 1005-1009.